\documentstyle[preprint,aps,pre,graphicx]{revtex}
\tightenlines
\begin{document}
\draft
\title{Sticky Spheres, Entropy barriers and Non-equilibrium phase transitions}
\author{S.L. Narasimhan$^{1*}$,  P.S.R. Krishna$^{2*}$ and K.P.N.Murthy$^{3+}$}
\address{$^1$High  Pressure  Physics Division, $^2$Solid State Physics Division,
\\ Bhabha Atomic Research Centre, Mumbai - 400 085, India}
\address{$^3$Materials  Science  Division, Indira Gandhi Centre for Atomic Research,
\\ Kalpakkam - 603 102, Tamilnadu, India}

\maketitle

\begin{abstract}
A   sticky   spheres   model  to  describe  slow  dynamics  of  a
non-equilibrium system is proposed. The dynamical slowing down is
due to the presence of entropy barriers. An  exact  steady  state
analysis  of the representative mean field equations, in the case
when the clusters  are  chosen  with  the  same   {$\grave{a}$ \it priori}
probability, demonstrates a non-equilibrium phase transition from
an exponential cluster size distribution to a powerlaw.

\end{abstract}


\section{Introduction}

When  a  macroscopic system in equilibrium at high temperature is
quenched rapidly to  low  temperature,  either  or  both  of  the
following  can happen. The system may get thermally arrested in a
metastable (local energy minimum) state much faster than it could
equilibrate  at  that  temperature  and  hence,  its   subsequent
dynamical  evolution  becomes  slow,  for details see [1]. On the
other hand, the system may still have thermal freedom to sample a
large number of equal or  almost  equal  energy  states,  upon  a
temperature quench, so that its dynamical evolution again becomes
slow,  for  example  see  [2].  In  other words, the system would
remain trapped for a long  time  due  to  the  presence  of  {\it
energy}   and/or   {\it  entropy}  barriers.  As  a  result,  the
relaxation  of  the  system  to  its  equilibrium  could   become
anomalously slow. It is often history dependent, usually referred
to  as  'aging', and could become progressively slower with time.
Glasses [3], obtained by the rapid quenching of liquids,  provide
simple  examples  of  aging  systems  which evolve slowly forever
towards their putative equilibrium states; granular systems whose
density  compaction  is  logarithmically  slow  in  response   to
mechanical   tapping  [4]  and  reaction-diffusion  systems  [5],
provide other recent examples. Quite often one finds  that  these
systems  develop  a  certain  degree of spatial disorder as well.
Experimental evidence for  such  a  scenario  has  recently  been
reported  in  the  literature [6]. An interesting problem in this
context is to see whether  simple  {\it  local}  dynamical  rules
could  be  devised so as to capture the essential features of the
non-equilibrium slow dynamics. In  particular,  it  would  be  of
interest  to  devise  dynamical  rules  that  could  lead to slow
logarithmic growth  of  length  scales  often  found  in  several
systems.  To  this  end, we propose in this paper a sticky sphere
model to describe the slow dynamics of a non-equilibrium system.

The  model  consists of hard spheres placed randomly on a regular
lattice. The energy of the system is defined in such a  way  that
nearest  neighbor  contacts between the spheres are energetically
favored, hence the name 'sticky' spheres. An  appropriate  length
scale  for  this  system  is  the  mean  cluster size. We present
numerical   evidence   to   show   that   this   quantity   grows
logarithmically  with  time  at  zero  temperature.  However, for
nonzero temperatures, it saturates asymptotically to a stationary
value. The model and the  simulation  details  are  discussed  in
section  II.  A  general  mean field formulation of this model is
presented in section III. An exact steady state analysis of  this
mean  field  model  for  the  special  case when the clusters are
chosen  with  the  same  {$\grave{a}$  \it  priori}  probability,
described  in  section  IV,  shows  a  phase  transition  from an
exponential to a power law cluster  size  distribution.  A  brief
summary of the results is presented in section V.

\section{Model}

Consider  a  regular  one  dimensional  lattice  of  size, $M+N$,
consisting of $N$ sites unoccupied and $M$ sites occupied by hard
spheres of size equal to  the  lattice  spacing.  Therefore,  the
spheres  on  nearest  neighbor  sites touch each other. We assume
periodic boundary conditions.

Let  us define the 'energy' of the system, $E(t)$, at time $t$ as
the negative of the total number of nearest neighbor contacts:

\begin{equation}
E(t) = - \sum _{k=1}^{M} (k-1)c_{k}(t)
\end{equation}

\noindent  where $c_{k}(t)$ is the total number of $k-mers$ ({\it
i.e.,} clusters consisting of $k$ spheres touching each other  at
time  $t$). We assume that the number of spheres in the system is
conserved: $\sum_{k=1}^{M} k\, c_{k}(t) = M$. The  lowest  energy
state  of  the system corresponds to having a single $M-mer$ with
energy, $E_{0} =  -(M-1)$,  and  may  henceforth  be  called  the
'ground  state'  of  the  system.  On the other hand, the highest
possible energy realizable for the system depends on  the  values
of both $M$ and $N$.

For  given  $M$  and  $N$,  we can always have a configuration of
spheres with a maximum of $(M-N)$ nearest neighbor contacts, with
$M\ >\ N$. This implies that the maximum energy  the  system  can
have  is  given  by,  $E_{max}(M,N)  = -(M-N)$. However, when the
system consists of only monomers,  which  can  be  realized  when
$M\le N$, the energy is zero. Thus we have,

\begin{equation}
E_{0} = -(M-1) ;\hskip.25in E_{max}(M,N) =
\left\{ \begin{array}{cc} -(M-N) & {\rm for} M > N \\
                                          0  & {\rm for} M \leq N \end{array} \right.
\end{equation}

\noindent  From  Eq. (1), it follows that the energy per particle,
$\epsilon  (t)  =   -1   +   [C(t)/M]$,   where   $C(t) =
\sum_{k=1}^{M}  c_{k}(t)$,  is  the  average number of clusters at
time $t$. Since $M/C(t)$ is just the mean cluster size,  $\Lambda
(t)$,  we  have  $\epsilon  (t)  =  -1  +  (1/\Lambda  (t))$,  or
equivalently, $\Lambda (t) = 1/[1+\epsilon (t)]$. Thus, we have a
'sticky' sphere system in which  nearest  neighbor  contacts  are
energetically favored.

We  start  from  an  initial  ($t=0$) configuration of the sticky
spheres placed on a one dimensional lattice segment in such a way
that the system is in the highest possible energy state for given
$M$ and $N$. At any instant of time $t$, we choose a $k-mer$ with
a pre assigned probability, $p_{k}$. Usually  we  take  $p_k$  as
$k/M$,  implying  thereby  that we choose a sphere at random with
the {$\grave{a}$ \it  priori}  probability,  $1/M$.  If  we  have
chosen  a  monomer ($k=1$), then it can hop either to its left or
to its right with equal probability. On the  other  hand,  if  we
have  chosen a $k-mer$ with $k>1$, then we choose one of its edge
spheres (or equivalently, edge particles) at  random  with  equal
probability. We call it the 'active' particle. We note that there
is  at  least one empty site available for the active particle to
hop. Consider the situation where we  have  chosen  the  leftmost
sphere of the $k-mer(k >1)$ as the active particle. This particle
can  hop  to the left. Let there be an $l-mer (l \geq 1)$ located
to the left of the active particle such that there are $n$  empty
sites in between them. If $n=1$, we simply move the particle into
the  available empty site because it does not cost energy. At the
end of this move,  we  have  an  $(l+1)-mer$  and  a  $(k-1)-mer$
separated  by  one  empty  site.  If  $n>1$,  then  we  have  two
possibilities  for  the  particle  to  hop,  as  illustrated   in
fig.1(a), and described below.

{\it  (i) Hopping to the nearest neighbor empty site:} If we move
the active particle to the nearest empty site, then we  would  be
creating  a  monomer  in the system. This process would therefore
cost one unit of energy. Hence, in order to  take  care  of  this
energy   cost,  we  move  it  to  the  nearest  empty  site  with
probability $e^{-\beta}$, where $\beta$ is  the  inverse  of  the
temperature.  At  the  end  of  this move, we will have a monomer
located in between an $l-mer$ and a $(k-1)-mer$.

{\it  (ii) Hopping to the farthest empty site:} If the above move
is not accepted, then we move the active particle to the farthest
empty site so that it sticks to the right edge  of  the  $l-mer$.
The  energy of the system (or equivalently, the number of nearest
neighbor contacts in the system) does not change. At the  end  of
this  move,  we  will have an $(l+1)-mer$ and a $(k-1)-mer$, with
$n$ empty sites in between them.

We  have  simulated  the above process for the case $M=N$ so that
the dynamics will cover the full range  of  energy  ($\epsilon  =
E/M$) from $0$ to $-1$. The {$\grave{a}$ \it priori} probability,
$p_{k}$,  for  choosing  a $k-mer$ is taken to be proportional to
$k$ in the simulation.  We  have  presented  in  Fig.2  the  mean
cluster size, $\Lambda (t)$, as a function of $ln(t)$ obtained by
averaging  the  data  over  $50$ independent runs for a system of
size  $N=16384$  and  temperatures  $\beta  =  2,4,5,6,8,10$  and
$\infty$.  We observe that $\Lambda (t)$ saturates asymptotically
for  temperatures   $T>0$,   whereas   it   continues   to   grow
logarithmically at $T=0$.

We   note   that   the  logarithmically  slow  dynamics  at  zero
temperature is purely due to  entropy  barriers  because  monomer
creation  is not possible at this temperature; the system evolves
only by the process of hopping to  the  farthest  neighbor  empty
site,  which  does  not  cost  energy.  In  this sense, our model
belongs to the same class of mean field models as that of  Ritort
[2]. In fact, we could anticipate this on heuristic grounds:

The  system  will  necessarily  have to be in the configurational
state consisting of a monomer and an $(N-1)-mer$ before it  might
be  able  to  reach the ground state by choosing the monomer with
probability $1/N$. This  is  a  rare  event  because  the  larger
cluster  will  always  loose  a  particle  with more probability.
Precisely the same situation prevails [7] in the  Ritort's  model
as  well.  Hence,  we  have  also  shown  in  Fig.2 the growth of
$\Lambda (t)$ obtained  from  Godr$\grave{e}$che-Luck  (GL)  mean
field formalism [8] of the Ritort's model as continuous lines. We
observe that our simulation data agree more or less with those of
the  GL values for asymptotic times $t>\tau _{GL}$, where we have
schematically shown the temperature dependence of $\tau _{GL}$ in
the inset of Fig.2. Clearly, $\tau _{GL} \rightarrow \infty $  as
$\beta  \rightarrow  \infty$,  and  the simulation data fall on a
line parellel to but below the GL line. The sticky sphere system,
therefore, admits of a mean field description  that  incorporates
the GL formalism at appropriate limits.

\section{Mean  field  formulation  of  a  one  dimensional sticky
sphere system}

The   hopping  of  a  single  particle  to  its  nearest/farthest
neighbour empty site can be incorporated easily in a  mean  field
description  by  considering  the dual representation obtained by
replacing particles by holes and holes by particles.  $k-mers  (k
\geq  1)$  of the sticky sphere system $S$ (Fig.1a) correspond to
empty intervals of length $k$ in its dual representation  $S^{*}$
(Fig.1b), and {\it vice versa}. The energy of the system is still
given by Eq.(1) except that $c_{k} (t)$ now stands for the number
of successive empty sites of length $k$ in $S^{*}$.

Consider  a  $k-mer$,  $K_{k}$, in $S$ having the empty intervals
$I_{m}$ and $I_{n}$ to its left and right respectively  (Fig.1a).
This  corresponds  to  the  empty  interval $I^{*}_{k}$ between an
$m-mer$, $K^{*}_{m}$, and an $n-mer$, $K^{*}_{n}$, in $S^{*}$. Let  $P$
be  the  rightmost particle of $K_{k}$. The hopping of $P$ to its
right nearest neighbour site, $Q$,  in  $S$  corresponds  to  the
dissociation  of the leftmost particle of $K^{*}_{n}$ in $S^{*}$. On
the other hand, hopping of $P$ to the  farthest  neighbour  site,
$R$  in  $S$  corresponds  to  the cluster $K^{*}_{n}$ moving as a
whole to the left  by  one  lattice  unit  in  $S^{*}$.  Thus,  the
nearest/farthest   neighbour   hopping   of  a  particle  in  $S$
corresponds  to  (single  particle)  dissociation/movement  of  a
cluster in $S^{*}$.

In  general, these processes may occur with probabilities $q_{1}$
and $q_{2}$ respectively. For convenience,  we  may  rescale  the
time   so  as  to  have  these  events  (namely,  single-particle
dissociation/movement of a cluster) occur with  the  rates  unity
and  $\omega  = q_{2}/q_{1}$ respectively. Spatial correlation in
the system may be ignored by treating the $k-mers (k \geq 1)$  in
$S^{*}$  as  point masses occupying single lattice sites only. This
leads to a simplified mean field description  of  the  system  in
terms  of a distribution of masses on $N$ lattice sites. We study
the stochastic evolution  of  the  system  in  the  thermodynamic
limit,  $M,\  N \rightarrow \infty $ with the mass density, $\rho
\equiv M/N$, remaining finite.

Let  $f_{k}(t)$ be the probability that a site will have mass $k$
at time $t$. By definition, $\sum  _{k=0}^{\infty}  f_{k}(t)  =1$
and  $\sum  _{k=0}^{\infty} kf_{k}(t) = \rho$. Let $p_{k}$ be the
{$\grave{a}$ \it priori} probability for choosing  a  $k-cluster$  and,  if
chosen,  let  $d_{k}$ be the {$\grave{a}$ \it priori} probability
for moving it by one lattice unit.  The  evolution  equation  for
$f_{k}(t)$ can now be written as

\begin{eqnarray}
\frac{df_{k \geq 2}(t)}{dt} = & \pi (t) f_{k-1}(t) - [\pi (t) + \lambda _{\beta}(t)p_{k}]f_{k}(t)
                                                 + \lambda_{\beta}(t)p_{k+1}f_{k+1}(t) \nonumber \\
        & - \omega \left\{ \left[ p_{k}d_{k} + \Delta(t) \right] f_{k}(t)
          -\sum_{n=1}^{k} p_{n}d_{n}f_{n}(t)f_{k-n}(t) \right\},
\end{eqnarray}

\noindent where,

\begin{eqnarray}
\pi (t) \equiv \sum _{n=1}^{\infty} p_{n}f_{n}(t) ;\hskip.25in \Delta (t) \equiv
                \sum _{n=1}^{\infty} p_{n}d_{n}f_{n}(t)  ; \nonumber \\
\lambda   _{\beta}(t)   \equiv  (1-e^{-\beta})s(t)  +  e^{-\beta}
;\hskip.25in  s(t) \equiv \sum _{n=1}^{\infty} f_{n}(t)
\end{eqnarray}

This  equation  consists  of  two parts, one corresponding to the
single particle dissociation and the other to the cluster  moving
by a lattice unit as a whole. Each part has both the gain and the
loss terms.

In  the  case of single particle dissociation, there are two gain
terms. The first one corresponds to the event  of  a  dissociated
particle   sticking   to   a   $(k-1)-cluster$.  The  second  one
corresponds to a particle dissociating  from  a  $(k+1)-cluster$,
taking  care  to  account  for  the  energy  cost,  $e^{-\beta}$,
involved in the event of its becoming a monomer.  Similarly,  the
first  of  the  loss  terms corresponds to a dissociated particle
sticking to a  $k-cluster$.  The  second  one  corresponds  to  a
particle  dissociating from a $k-cluster$, taking care to account
for the energy cost, $e^{-\beta}$, in the event of its becoming a
monomer. The probability of choosing a $k-cluster$, $p_{k}(k \geq
1)$, has been introduced appropriately.

In  the  case of a cluster moving by one lattice unit as a whole,
the gain term corresponds to an $n-cluster (1  \leq  n  \leq  k)$
coming  to stick to a $(k-n)-cluster$. The event of a $k-cluster$
moving out as well as that of a cluster coming in to stick  to  a
$k-cluster$  constitute the loss terms. The probability of moving
a  cluster,  $p_{n}d_{n}(n  \geq   1)$,   has   been   introduced
appropriately.

Similarly,  the  master  equations  satisfied  by  the fractions,
$f_{0}(t)$ and $f_{1}(t)$, can be written as follows:

\begin{eqnarray}
\frac{df_{1}(t)}{dt} = & \mu _{\beta} (t) f_{0}(t) - [\pi (t) + p_{1}]f_{1}(t)
                                                    + \lambda_{\beta}(t)p_{2}f_{2}(t) \nonumber \\
        & - \omega \left\{ [p_{1}d_{1} + \Delta(t)] f_{1}(t) - p_{1}d_{1}f_{1}(t)f_{0}(t) \right\},
\end{eqnarray}
\begin{equation}
\frac{df_{0}(t)}{dt} =  -\mu _{\beta}(t) f_{0}(t)  + p_{1}f_{1}(t)
                                                    + \omega s(t) \Delta (t),
\end{equation}

\noindent      where,      $      \mu      _{\beta}(t)     \equiv
p_{1}(1-e^{-\beta})f_{1}(t) + \pi (t)e^{-\beta}$. In this  model,
the parameters  $\omega$ and  $d$'s  are all
assumed  to  be  temperature  independent.

The  mean  field  equations,  obtained  by  ignoring  the spatial
extensions of  $k-mers  (k  \geq  1)$  in  $S^{*}$,  provide  the
simplest  representation of the nearest/farthest neighbour single
particle hopping of a sticky sphere system $S$. Yet, we  can  not
assume {$\grave{a}$ \it priori} that they describe the asymptotic
dynamical behaviour of $S$. Because, the probability, $p_{k}$, of
choosing  a  $k-mer$  actually  stands  for  the  probability  of
choosing the empty interval bounded on one side by the $k-mer$ of
interes. It is also  important  to  note  that  the  presence  or
absence  of  the  aggregation term, $F_{k} \equiv \sum _{n=1}^{k}
p_{n}d_{n}f_{n}f_{k-n}$, in  Eq.3  corresponds  to  the  specific
monomer  dynamics  implemented  in  $S$, {\it viz}., whether they
jump to their  farthest  or  to  their  nearest  neighbour  sites
respectively.  However,  in the case when the clusters are chosen
with equal {$\grave{a}$ \it priori} probability, empty  intervals
are   also   chosen   with  the  same  {$\grave{a}$  \it  priori}
probability (say, $p_{k}=1$); hence,  the  mean  field  equations
could  provide an adequate description of $S$. Moreover, it turns
out that an exact steady state analysis of these equations can be
carried out in this case.

\section{Steady  state  analysis}

Here,  we  consider  the  case  $p_{j} = d_{j} = 1$, for which an
exact steady state analysis can be carried out. It is clear  from
Eq.(3-6)  that  the  generating  function, $Q_{\beta}(z,t) \equiv
\sum  _{k=1}^{\infty}  z^{k}f_{k}(t)$,  satisfies  the  following
equation:

\begin{equation}
\frac      {\partial      Q_{\beta}(z,t)}{\partial      t}      =
Q_{\beta}^{2}(z)-\left[ a(z) + \frac {b}{z} \right]Q_{\beta} + c(z)
\end{equation}

\noindent where,

\begin{eqnarray}
a(z) & = & 2s + \frac {s}{\omega} +\frac {\lambda_{\beta}}{\omega} -\frac {sz}{\omega} \nonumber\\
b(z) & = & -\frac {\lambda _{\beta}}{\omega} \\
c(z) & = & \lambda _{\beta} (z-1)\left[ \frac {\mu _{\beta} (1-s)}{\omega} -s^{2} \right] + zs^{2} \nonumber
\end{eqnarray}

\noindent  In  order  to  study  the steady state behavior of the
system, we set $\partial Q_{\beta}(z,t)/\partial t =0$ and choose
the  root  of  the   resulting   quadratic   equation   so   that
$Q_{\beta}(z=0) = 0$ is ensured:

\begin{equation}
2Q_{\beta}(z) = [a(z) + \frac {b}{z}] -
\sqrt{\left[ a(z) + \frac {b}{z} \right]^{2} - 4c(z)}.
\end{equation}

\noindent Simplifying the algebra, we can show that

\begin{equation}
\left[ a(z) + \frac {b}{z} \right]^{2} - 4c(z) =\left\{\frac {s(z-1)}{\omega z}
                                                      \right\}^{2}(z-z_{1})(z-z_{2}),
\end{equation}

\noindent where, the roots, $z_{1,2}$ are given by

\begin{equation}
z_{1,2} = \left( \frac {\tau }{s} + 1 - \tau \right) \left[ 1 + 2 \omega \mp 2
               \sqrt{ \omega ^{2} + \omega } \right] ;
\hskip.25in  \tau \equiv e^{-\beta}.
\end{equation}

\noindent Hence, we have the generating function,
\begin{equation}
Q_{\beta}(z) = \frac {2 \omega s + s + \lambda _{\beta}}{2 \omega} - \frac {\lambda_{\beta}}{2 \omega z}
                           - \frac {sz}{2 \omega} +\frac {s(1-z)}{2 \omega z} \sqrt{(z-z_{1})(z-z_{2})}
\end{equation}

\noindent  The  value  of  $s$  is  fixed  by the conservation of
particle density, $\rho$:

\begin{equation}
\rho = \left\{ \frac {\partial Q_{\beta}(z)}{\partial z} \right\} _{z=1} = \frac {1}{2 \omega}
                             \left[\lambda _{\beta} - s\left(1 + \sqrt{(1-z_{1})(1-z_{2})}  \right) \right]
\end{equation}

\noindent  For  a  given  $\omega$, it is clear that the value of
$z_{1}$, being always less than $z_{2}$, should not be less than
unity for $\rho$ to be real.  As  $\rho$  increases,  the  steady
state  value  for the number of clusters, $s$, increases, thereby
reducing the values of $z_{1,2}$. Hence, we have the condition,

\begin{equation}
s \leq \frac {\tau P_{1}(\omega)}{1-(1-\tau)P_{1}(\omega)} ;\hskip.25in  P_{1}(\omega) \equiv
                         1+ 2\omega -2\omega \sqrt{1+ \frac{1}{\omega}}
\end{equation}

\noindent The equality sign defines the critical value $s_{c}$ at
which the root $z=1$, and hence the critical density,

\begin{equation}
\rho _{c} = \frac {\tau \rho _{c}^{0}}{\tau + 2\omega (1-\tau) \rho _{c}^{0}} ;\hskip.25in
                   \rho _{c}^{0} \equiv \sqrt{ 1 +\frac {1}{\omega} }-1
\end{equation}

\noindent The number of clusters will not increase beyond $s_{c}$
for  $\rho  >  \rho  _{c}$.  It  is  of  interest to consider the
question of how  this  inequality  influences  the  cluster  size
distribution.  To  this  end,  we  consider the following contour
integral,

\begin{equation}
f_{k} = \oint \frac {Q_{\beta}(z)}{z^{k+1}} dz.
\end{equation}

\noindent The contour is chosen suitably so that only the portion
of   the  contour  above  and  below  the  branch  cut  $z=z_{1}$
contributes to the integral. The number of $k-mers,  f_{k},$  has
the asymptotic exponential form, $(1/z_{1})^{k}$ for $\rho < \rho
_{c}$  whereas  it  has  a power law form, $k^{-5/2}$ for $\rho =
\rho _{c}$; as the density is  increased  beyond  $\rho_{c}$,  in
addition  to  the  power  law  decay, the distribution develops a
delta function peak corresponding to an 'infinite' aggregate.

However,  at zero temperature, $z_{1} < 1$ for all nonzero values
of $\omega$; therefore, the above steady  state  analysis  breaks
down.  In  fact,  the  condition  expressed  by  Eq.  (14) can be
rewritten as

\begin{equation}
\tau \geq \tau _{c} ;\hskip.25in  \tau _{c} \equiv \frac {s[1-P_{1}(\omega)]}{(1-s)P_{1}(\omega)}
\end{equation}

\noindent  For  a  given  $\omega  $,the  value  of  $\tau  _{c}$
increases as we increase the particle density, until  it  becomes
equal  the  given  temperature;  beyond  this,  the  steady state
analysis breaks down. In other  words,  the  steady  state  phase
transition  from  the  'exponential'  regime to the 'aggregating'
regime is observable only  in  a  limited  range  of  temperature
decided  by $\omega$ and $\rho$. The infinite temperature version
of a related model has been discussed by Majumdar,  Krishnamurthy
and Barma [9].

\section{Summary}

In  this  paper,  we have presented a generic sticky sphere model
for describing the non-equilibrium  behavior  of  a  system  fast
quenched  to  a  low  temperature. The evolution of the system is
based on a {\it local}  dynamical  rule  -  the  nearest/farthest
neighbor  hopping of a randomly chosen particle. The mean cluster
size, defining a length  scale  for  the  system,  asymptotically
saturates  to a stationary value at nonzero temperatures, whereas
it grows logarithmically with time at zero temperature.  We  have
presented  a  general  mean  field  formulation of this model and
solved it exactly for the case when the clusters are  chosen  and
moved with the same {$\grave{a}$ \it priori} probability. We have
shown that the steady state cluster size distribution undergoes a
phase  transition  (in  appropriate  temperature  range)  from an
exponential form to a power law with an additional delta function
peak corresponding to an 'infinite' cluster.

One of the authors (S.L.N) thanks Y. S. Mayya and Amitabh Joshi
for fruitful discussions.

\noindent $^*$glass@apsara.barc.ernet.in;\\
$^+$kpn@igcar.ernet.in

\begin{figure}
\includegraphics[width=6in]{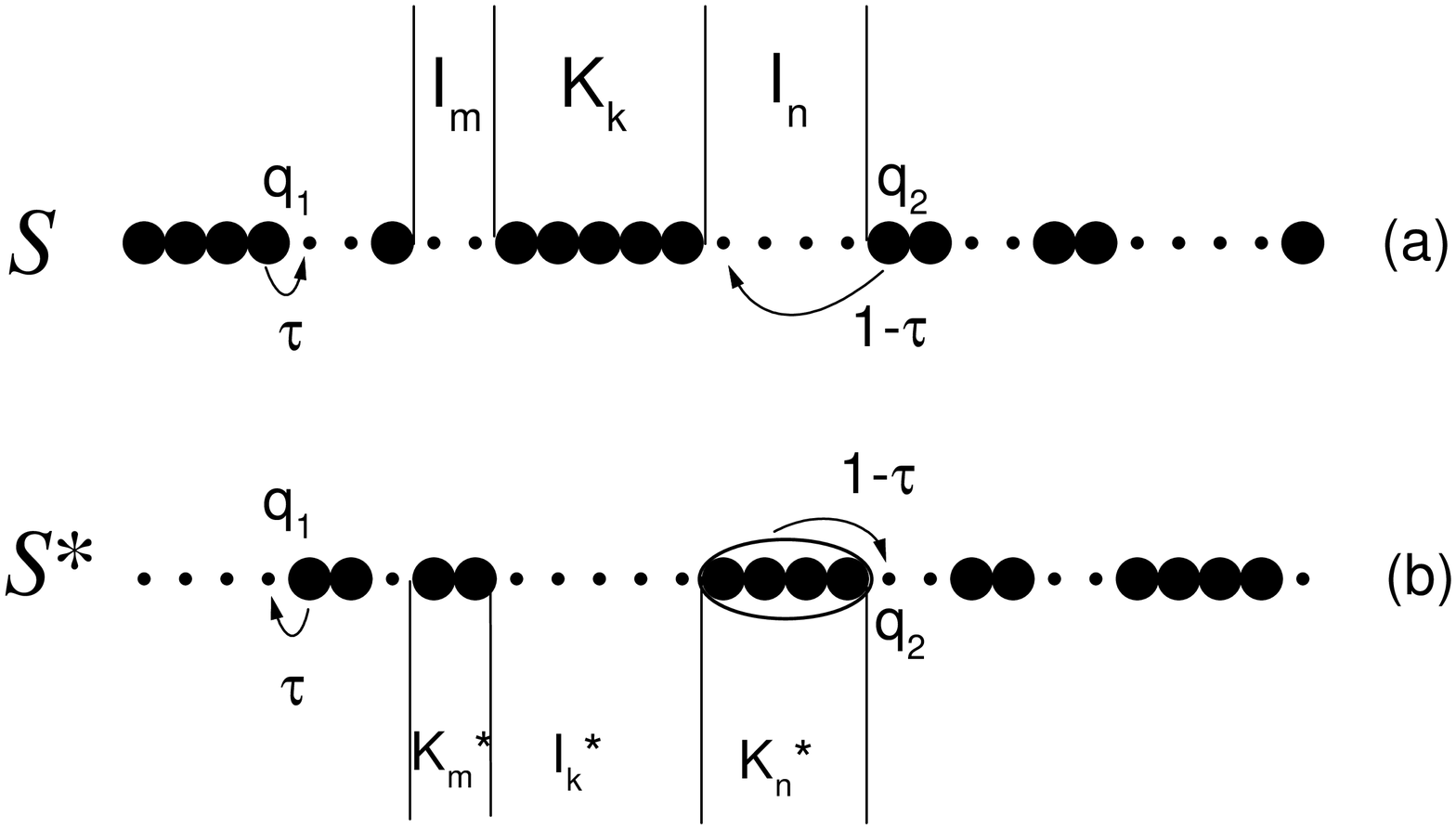}

\caption{(a)Sticky  sphere  model:  the dynamical moves available
for  a  chosen  particle;  (b)  its  dual   representation:   the
corresponding dynamical moves. $\tau \equiv e^{-\beta}$.}

\end{figure}

\begin{figure}
\includegraphics[width=6in]{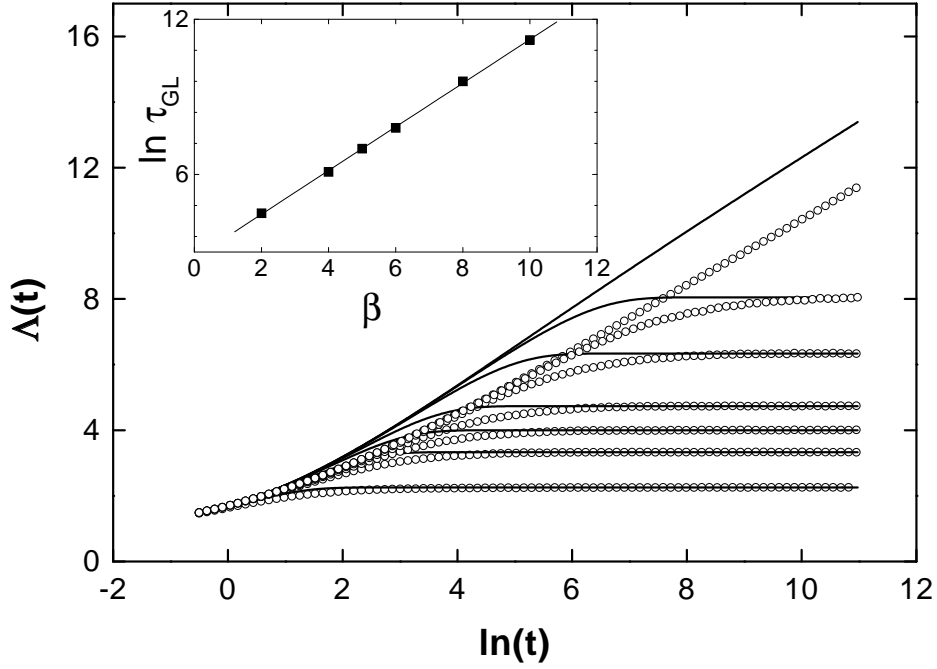}

\caption{Mean Cluster size for $N=16384$.
The   time  $t$  is  measured  in  units  of  $1/N$.  Inverse  of
temperature $\beta = \infty, 10,8,6,5,4$ and  $2$,  from  top  to
bottom. Open circles represent simulation data obtained as $50$ runs
averages;   Continuous   lines   have    been    obtained    from
Godr$\grave{e}$che  and  Luck  mean  field  formalism[4]  of  the
Ritort's  model.  {\it  Inset}:  Temperature  dependence  of
$\tau_{GL}$, the times beyond which simulation more or less agrees with GL.}

\end{figure}

\end{document}